\documentclass[12pt]{article}
\usepackage{amsmath,epsfig,amssymb,bbm}
\usepackage{cite}

\begin{document}

\title{\vbox{
\baselineskip 14pt
\hfill \hbox{\normalsize WU-HEP-10-02} \\
\hfill \hbox{\normalsize KUNS-2300}\\
\hfill \hbox{\normalsize YITP-10-85} } \vskip 2cm
\bf Non-Abelian Discrete Flavor \\ Symmetries on Orbifolds \vskip 0.5cm
}
\author{Hiroyuki~Abe$^{1,}$\footnote{email:
 abe@waseda.jp}, \
Kang-Sin~Choi$^{2,}$\footnote{email:
  kschoi@gauge.scphys.kyoto-u.ac.jp}, \
Tatsuo~Kobayashi$^{2,}$\footnote{
email: kobayash@gauge.scphys.kyoto-u.ac.jp} \ \\ 
Hiroshi~Ohki$^{3,}$\footnote{email: ohki@kmi.nagoya-u.ac.jp
}  \ and \
Manabu~Sakai$^{4,}$\footnote{email: msakai@yukawa.kyoto-u.ac.jp}
\\*[20pt]
$^1${\it \normalsize
Department of Physics, Waseda University, Tokyo 169-8555, Japan} \\
$^2${\it \normalsize
Department of Physics, Kyoto University,
Kyoto 606-8502, Japan} \\
$^3${\it \normalsize Kobayashi-Maskawa Insitute for the 
Origin of Particles and the Universe (KMI)} \\ { \it Nagoya University, 
  Nagoya 464-8602, Japan} \\
$^4${\it \normalsize 
Yukawa Institute for Theoretical Physics, Kyoto University, 
Kyoto 606-8502, Japan}
}

\date{}

\maketitle
\thispagestyle{empty}

\begin{abstract}
We study non-Abelian flavor symmetries 
on orbifolds, $S^1/Z_2$ and $T^2/Z_3$.
Our extra dimensional models realize 
$D_N$, $\Sigma(2N^2)$, $\Delta(3N^2)$ and $\Delta(6N^2)$ including 
$A_4$ and $S_4$.
In addition, one can also realize their subgroups such as 
$Q_N$, $T_7$, etc.
The $S_3$ flavor symmetry can be realized on both 
$S^1/Z_2$ and $T^2/Z_3$ orbifolds.
\end{abstract}

\newpage

\section{Introduction}

Non-Abelian discrete flavor symmetries 
play an important role in particle physics.
In particular, several non-Abelian discrete flavor 
symmetries such as $S_3$, $S_4$, $A_4$, 
$D_N$, $Q_N$, $T_7$, $\Delta(27)$ and $\Delta(54)$,  
 have been used to derive lepton mass 
matrices with the large mixing angles as well as 
quark mass matrices.
(See e.g \cite{ref:S3,ref:S4,Altarelli:2005yp,ref:A4,ref:DN,ref:QN,
Ma:2007ia,Hagedorn:2008bc,Luhn:2007uq,ref:Delta27,Escobar:2008vc,
Ishimori:2008uc} 
and also see 
for reviews \cite{Altarelli:2010gt,Ishimori:2010au}.)
In addition to quark/lepton mass matrices, 
non-Abelian discrete flavor symmetries are also important 
to control supersymmetry (SUSY) breaking terms such as 
squark/slepton masses and A-terms.
(See e.g. \cite{Hall:1995es,Ko:2007dz}.)

Some of non-Abelian discrete symmetries are 
symmetries of geometrical solids.
Hence, we expect that non-Abelian discrete flavor symmetries 
would be originated from extra dimensional theories 
and superstring theories.
Indeed, orbifolds have certain geometrical symmetries.
Thus, field theories on orbifolds can realize 
non-Abelian discrete flavor symmetries and 
localized modes on fixed points of orbifolds 
correspond to certain reducible/irreducible representations 
\cite{Altarelli:2006kg,Adulpravitchai:2009id}.

In general, it is not possible to realize desired nontrivial 
non-Abelian groups and representations purely by geometrical
symmetry of the orbifold. 
However there is another room that we can assign different $Z_N$
charges to the fields of different elements in a multiplet. 
This results in enhanced non-Abelian discrete symmetries. 
Thus our classification is characterized as the following properties:
\begin{enumerate}
\item We specify the geometric distribution. 
On special locations, like the fixed points of toroidal orbifolds, 
fields are democratically distributed. 
They would have the symmetry of the solid or 
in general a pernutation symmetry $S_N$ in the low energy Lagrangian. 
\item Since fields are complex valued, 
we collect such fields on fixed points to form multiplets 
and allow complex representation, 
which is not necessary irreducible.
\end{enumerate}
With these, we study how far the symmetry can go. 
The second assignment easily mimics the symmetry from stringy selection
rules. 
For example, localized (twisted) strings at orbifold fixed points have definite 
$Z_N$ charges, which control the selection rules of 
allowed couplings.
Combinations between such $Z_N$ symmetries and 
geometrical symmetries enhance flavor symmetries.
Then, in heterotic string orbifold models 
we can realize non-Abelian discrete flavor symmetries larger 
than geometrical symmetries of orbifolds 
such as $D_4$ and $\Delta(54)$ 
\cite{Kobayashi:2004ya,Kobayashi:2006wq,Ko:2007dz}.\footnote{
Similar symmetries can be realized in 
magnetized/intersecting brane models 
\cite{Abe:2009vi}.}

It is important to extend the above analysis.
We study field theories on orbifolds.
Inspired by heterotic string orbifold models, 
we assume $Z_N$ charges for modes localized
on fixed points.
Only limited  kinds of $Z_N$ charges can be realized in 
heterotic string orbifold models.
That is the reason why there are a strong restriction 
on realization of  non-Abelian discrete flavor symmetries 
in heterotic orbifold models.
However, here we consider localized modes 
with generic $Z_N$ charges.
Those localized modes would have symmetries, 
which are combinations between $Z_N$ symmetries and 
geometrical symmetries of orbifolds.
Then, those localized modes correspond to 
definite reducible/irreducible representations, 
while the bulk modes would correspond to 
(trivial) singlets in our theory like heterotic orbifold 
models.

We are interested in three-generation models.
Thus, we consider orbifolds with two or three fixed points, 
because the number of fixed points corresponds to 
the number of generations.
For the orbifold with two fixed points, 
we expect one generation may originate, e.g. from a bulk field.
Hence, we restrict ourselves to $S^1/Z_2$ and $T^2/Z_3$ 
orbifolds, which have two and three fixed points, respectively.
(See for geometrical aspects of other orbifolds, 
e.g.\cite{Dixon:1985jw}.)
The $T^2/Z_4$ orbifold also has two fixed points of the same deficit
angles, which has the $Z_2$ permutation symmetry,  
and would lead to 
the same results as ones on the $S^1/Z_2$ orbifold.
Also note that the orbifold with a single fixed point 
does not lead to non-Abelian discrete flavor symmetries 
in our approach.
Thus, we do not consider such orbifolds like, $T^2/Z_6$.
Using $S^1/Z_2$ and $T^2/Z_3$ orbifolds, 
we study which flavor symmetries can appear 
and which representations can be realized.

This paper is organized as follows.
In section 2 and 3, we consider $S^1/Z_2$ and $T^2/Z_3$ 
orbifolds, respectively, 
and study which flavor symmetries appear 
and which representations can realized.
Section 4 is devoted to conclusion and discussion.

\section{$S^1/Z_2$ orbifold}

Here, we consider the $S^1/Z_2$ orbifold, which 
has two fixed points.
We denote fields localized at these two fixed points 
by $\phi_1$ and $\phi_2$.
The orbifold has the geometrical $Z_2$ symmetry, which 
permutes these two fixed points.
Such a permutation symmetry acts on two localized fields as 
\begin{eqnarray}
B= \left(
\begin{array}{cc}
0 & 1 \\
1 & 0
\end{array}
\right),
\end{eqnarray}
on the basis $(\phi_1,\phi_2)^T$.

Here, we assume the Abelian $Z_N$ symmetry and 
the fields, $\phi_1$ and $\phi_2$, have $Z_N$ charges, $q_1$ and $q_2$.
That is, such a $Z_N$ symmetry is represented as 
\begin{eqnarray}\label{eq:A-DN}
A= \left(
\begin{array}{cc}
\rho^{q_1} & 0 \\
0 & \rho^{q_2}
\end{array}
\right),
\end{eqnarray}
on the basis $(\phi_1,\phi_2)^T$,
where $\rho = \exp(2\pi i/N)$.

The full flavor symmetry is a closed algebra including 
the generators $A$ and $B$.
Such a symmetry would become non-Abelian symmetry, 
although each of the geometrical symmetry $B$ and 
the $Z_N$ symmetry $A$ corresponds to an Abelian symmetry.  
For example, when $\rho^{q_1}=1$ and $\rho^{q_2}=-1$, 
such a symmetry is realized in heterotic string on 
the $S^1/Z_2$ orbifold and that corresponds to 
the $D_4$ symmetry \cite{Kobayashi:2004ya,Kobayashi:2006wq,Ko:2007dz}.
The two fields $(\phi_1,\phi_2)^T$ correspond to a $D_4$ doublet 
${\bf 2}$.
Here we consider more generic values of $Z_N$ charges, $q_i$.
We can think of different signs in each element of $B$, 
but they would be redundant.

\subsection{$S_3$ and $D_N$}

First, let us consider the model with $q_1=-q_2$ (mod $N$), i.e.
\begin{eqnarray}
A= \left(
\begin{array}{cc}
\rho' & 0 \\
0 & \rho'^{-1}
\end{array}
\right),
\end{eqnarray}
where $\rho' = \exp(2\pi iq_1/N)$.
The generators $A$ and $B$ satisfy 
\begin{eqnarray}
B^2=I, \qquad BAB=A^{-1},
\end{eqnarray}
where $I$ denotes the identity.
Furthermore, when $q_1=1$, we have 
\begin{eqnarray}
A^N=I. 
\end{eqnarray}
Their closed algebra is $D_N$, which has $2N$ elements.
All of $D_N$ elements are written as $A^mB^k$ with 
$m=0,\cdots,N-1$ and $k=0,1$.
When $N=$ odd, the $D_N$ group has 
two singlets ${\bf 1}_{\pm}$ and $(N-1)/2$ doublets 
${\bf 2}_k$ ($k=1,\cdots,(N-1)/2$) as irreducible 
representations.
When $N=$ even, there are 
four singlets  ${\bf 1}_{\pm \pm}$ and $(N/2-1)$
doublets, ${\bf 2}_k$ ($k=1,\cdots,(N/2-1)$).

When $N=$ odd, the localized fields $(\phi_1,\phi_2)$ 
with $Z_N$ charges $(k,-k)$ correspond to the 
doublets ${\bf 2}_k$, where $k\neq 0$ (mod $N$).
When the localized fields $(\phi_1,\phi_2)$ have 
trivial $Z_N$ charges, i.e. $k=0$ (mod $N$), 
the generator $A$ is represented trivially, i.e. $A=I$.
Then, we can use the basis $(\phi_1+\phi_2,\phi_1-\phi_2)$ 
such that the generator $B$ is diagonalized as 
diag$(1,-1)$.
That is, these fields correspond to two singlets, 
${\bf 1}_+$ and ${\bf 1}_-$.
Note that the combination of ${\bf 1}_+$ and ${\bf 1}_-$ appears 
by the fields localized on two fixed points, but 
 one of them can not appear without the other singlet.
These results are shown in Table \ref{tab:DN-odd}.

\begin{table}[t]
\begin{center}
\begin{tabular}{|c|c|} \hline
$Z_N$ charges of $(\phi_1,\phi_2)$ & Representation of $D_N$ \\ \hline \hline
$(k,-k)$ (mod $N$)& ${\bf 2}_k$ \\
 $k\neq 0$ &  \\ \hline 
$(0,0)$ & ${\bf 1}_{+} + {\bf 1}_{-}$ \\ \hline
\end{tabular}
\end{center}
\caption{$D_{N={\rm odd}}$ representations of the fields 
localized on the
  $S^1/Z_2$ orbifold.}
\label{tab:DN-odd}
\end{table}

Note that $S_3$ is isomorphic to $D_3$.
Thus, the $S_3$ flavor symmetry is included in the above 
analysis and the localized fields with the $Z_3$ charges 
$(1,2)$ and $(0,0)$ on two fixed points 
correspond to the doublet ${\bf 2}$ and the combination 
${\bf 1}_+ + {\bf 1}_-$, respectively, where the 
generator $A$ is represented as  
$A=$ diag$(\omega,\omega^2)$ and $A=$ diag$(1,1)$ 
with $\omega=\exp (2 \pi i/3)$.
In the next section, we study another type of realization of 
$S_3$ on the $T^2/Z_3$ orbifold, where 
the fields localized on three fixed points of $T^2/Z_3$ 
correspond to the combination, e.g. ${\bf 1}_+ + {\bf 2}$.
Thus, we can realize the $S_3$ flavor symmetry on both 
$S^1/Z_2$ and $T^2/Z_3$, but 
combinations of representations appearing on fixed points 
are different from each other.

Similarly, we can study the model with $N=$ even.
The localized fields $(\phi_1,\phi_2)$ 
with $Z_N$ charges $(k,-k)$ correspond to the 
doublets ${\bf 2}_k$, where $k\neq 0$ (mod $N/2$).
The fields with $k=0$ (mod $N$) has the same situation as 
those for $N=$ odd in the above case.
In addition, when $k=N/2$, the generator $A$ is represented 
as $A=-I$.
Thus, we can use the field basis such that the generator 
$B$ is diagonalized as diag$(1,-1)$.
As a result, the  localized fields $(\phi_1,\phi_2)$ with 
the trivial $Z_N$ charges $k=0$ correspond to 
${\bf 1}_{++}$ and ${\bf 1}_{--}$.
On the other hand, the  localized fields $(\phi_1,\phi_2)$ with 
the $Z_N$ charges $k=N/2$ (mod $N$) correspond to 
${\bf 1}_{+-}$ and ${\bf 1}_{-+}$.
Here, signs in the subscripts mean as follows.
The first sign denotes the eigenvalue of $B$ and 
the second sign denotes the eigenvalues of $AB$.
Note that only the combination ${\bf 1}_{++} + {\bf 1}_{--}$ or 
${\bf 1}_{+-} +{\bf 1}_{-+}$ appears as the fields localized
on two fixed points, but other combinations or only one singlet 
can not appear.
These results are shown in Table \ref{tab:DN-even}.

\begin{table}[t]
\begin{center}
\begin{tabular}{|c|c|} \hline
$Z_N$ charges of $(\phi_1,\phi_2)$ & Representation of $D_N$ \\ \hline \hline
$(k,-k)$ & ${\bf 2}_k$ \\
 $k\neq0$ (mod $N/2$) &  \\ \hline 
$(0,0)$ & ${\bf 1}_{++} + {\bf 1}_{--}$ \\ \hline
$(N/2,N/2)$ (mod $N$) &  ${\bf 1}_{+-} + {\bf 1}_{-+}$ \\ \hline
\end{tabular}
\end{center}
\caption{$D_{N={\rm even}}$ representations of the fields
  localized on the
  $S^1/Z_2$ orbifold.}
\label{tab:DN-even}
\end{table}

Here, we comment on bulk fields.
The generator $B$ can be identified as the $Z_2$ reflection 
of $S^1/Z_2$.
In general, bulk fields, e.g. vector and spinor fields, have 
$Z_2$ even and odd zero-modes.
Those $Z_2$ even and odd zero-modes would correspond to 
a trivial singlet and a non-trivial singlet with $B=-1$.
However, $Z_2$ odd zero-modes have no direct couplings with 
localized modes, while $Z_2$ even zero-modes can couple.
Hence, among bulk modes, only trivial singlet fields 
would be useful in 4D effective field theory, 
at least in simple models.\footnote{
Non-trivial singlets of bulk fields may play an important role  
in complicated models, but here we do not consider such a possibility 
further.}
At any rate, a trivial singlet appears as a bulk field.
For example, for $D_N$ with $N=$ odd, 
we have claimed that whenever singlets appear on two fixed points, 
they always appear as the combination ${\bf 1}_+ + {\bf 1}_-$.
Here we can write mass terms between ${\bf 1}_+$ of 
the bulk and localized modes.
Then, below such a mass scale, the light modes include only 
the trivial singlet ${\bf 1}_-$.
This implies that any combinations of ${\bf 1}_+$, ${\bf 1}_-$ and 
${\bf 2}$ can appear on the $S^1/Z_2$ orbifold.
Hence, we can realize any $S_3$ models and $D_N$ models with 
$N=$ odd on the $S^1/Z_2$ orbifold.

A similar comment is also available for $D_N$ with $N=$ even.
That is, the bulk fields, which would play a role, correspond 
to trivial singlets, and such trivial singlets have 
$D_N$-invariant  mass terms with trivial singlets on the fixed points. 
Also these comments would be applicable to other 
flavor symmetries, which will be discussed 
in the following sections.

\subsection{$\Sigma(2N^2)$}

Next, let us consider more generic assignment of 
$Z_N$ chargers, i.e. $q_1\neq -q_2$ (mod $N$).
In this case, the closed algebra corresponds to 
$\Sigma(2N^2)$, which is isomorphic to 
$(Z_N \times Z_N) \rtimes Z_2$ \cite{Ma:2007ia,Ishimori:2010au}.
Two $Z_N$ generators $A$ and $A'$ can be written by 
\begin{eqnarray}
A=\left(
\begin{array}{cc}
1 & 0 \\
0 & \rho
\end{array}
\right), \qquad 
A'=\left(
\begin{array}{cc}
\rho & 0 \\
0 & 1
\end{array}
\right),
\end{eqnarray}
where $\rho = \exp (2\pi i/N)$.
Then, the generators, $A$, $A'$ and $B$, satisfy 
\begin{eqnarray}\label{eq:Sigma-algebra}
A^N=A'^N=B^2=I, \qquad AA'=A'A, \qquad 
BAB=A'.
\end{eqnarray}
All of $\Sigma(2N^2)$ elements can be written by 
$B^kA^mA'^n$ for $k=0,1$ and $m,n=0,1,\cdots,N-1$.
Explicitly, all of $\Sigma(2N^2)$ elements are 
represented by 
\begin{eqnarray}
\left(
\begin{array}{cc}
\rho^m  & 0 \\
0 & \rho^n 
\end{array}
\right), \qquad 
\left(
\begin{array}{cc}
0 & \rho^m  \\
 \rho^n  & 0
\end{array}
\right).
\end{eqnarray}
Note that $\Sigma(2N^2)$ for $N=2$ is isomorphic to 
$D_4$.

The localized fields $(\phi_1,\phi_2)$ with $Z_N$ charges 
$(m,n)$ for $m\neq n$ correspond to doublets of $\Sigma(2N^2)$, 
${\bf 2}_{m,n}$.
On the other hand, when the localized fields $(\phi_1,\phi_2)$ 
have $Z_N$ charges, $(m,m)$, the $Z_N$ generator can 
be represented by $\rho^nI$.
For these fields, we can use the field basis diagonalizing 
the generator $B$, i.e. $\phi_1 \pm \phi_2$, where 
the generator $B$ is represented by $B=\pm 1$.
Thus, these fields correspond to the combination of 
two singlets, ${\bf 1}_{+m}+ {\bf 1}_{-m}$.
These results are shown in Table \ref{tab:Sigma}.

\begin{table}[t]
\begin{center}
\begin{tabular}{|c|c|} \hline
$Z_N$ charges of $(\phi_1,\phi_2)$ & Representation of $\Sigma(2N^2)$ \\ \hline \hline
$(m,n)$ & ${\bf 2}_{m,n}$ \\
 $m\neq n$  &  \\ \hline 
$(m, m)$ & ${\bf 1}_{+m} + {\bf 1}_{-m}$ \\ \hline
\end{tabular}
\end{center}
\caption{$\Sigma(2N^2)$ representations of the  fields localized on the
  $S^1/Z_2$ orbifold.}
\label{tab:Sigma}
\end{table}

The $\Sigma(2N^2)$ groups include several subgroups 
and such subgroups can also be realized on the $S^1/Z_2$ orbifold.
One subgroup of $\Sigma(2N^2)$ is the $D_N$ group, 
and its realization on the $S^1/Z_2$ orbifold has been 
studied in the previous section.
Another non-trivial subgroup of $\Sigma(2N^2)$ for $N=$ even 
is the $Q_N$ group.
We define $\tilde A=A^{-1}A'$ and 
$\tilde B =BA'^{N/2}$.
By use of the algebra (\ref{eq:Sigma-algebra}), 
it is found that 
they satisfy the following algebraic relations,
\begin{eqnarray}
\tilde A^N=I, \qquad \tilde B^2 = \tilde A^{N/2}, \qquad 
\tilde B^{-1}\tilde A \tilde B = \tilde A^{-1}.
\end{eqnarray}
These are the algebraic relations of $Q_N$ generators.
Indeed, the closed algebra of $\tilde A$ and $\tilde B$ 
corresponds to $Q_N$ and 
all of $Q_N$ elements can be written by $\tilde A^m \tilde B^k$ 
with $m=0,1\cdots,N-1$ and $k=0,1$.
Thus, a proper breaking of $\Sigma(2N^2)$ would lead to 
the $Q_N$ flavor symmetry \cite{Ishimori:2010au}.
For example, a vacuum expectation value of 
the singlet scalar ${\bf 1}_{-m}$ with $m=$ odd  
would break the symmetries generated by 
 $B$ and $A'$, but the above $Q_N$ symmetry 
would remain.
Then, the doublet ${\bf 2}_{m+k,m}$ of $\Sigma(2N^2)$ 
becomes ${\bf 2}_{k}$ of $Q_N$, while 
the doublet ${\bf 2}_{m+N/2,m}$ of $\Sigma(2N^2)$  decomposes 
to two singlets ${\bf 1}_{+-} + {\bf 1}_{-+}$ in 
$Q_N$.
Furthermore, the singlets, ${\bf 1}_{+m}$ and 
${\bf 1}_{-m}$, of $\Sigma(2N^2)$ correspond to 
the $Q_N$ singlets, ${\bf 1}_{--}$ and ${\bf 1}_{++}$, 
respectively.
These results are summarized in Table \ref{tab:QN}.

Similarly, we could obtain other subgroups, e.g. 
$(Z_4 \times Z_2) \rtimes Z_2$, which is a subgroup of 
$\Sigma(32)$ \cite{Ishimori:2010au}.

\begin{table}[t]
\begin{center}
\begin{tabular}{|ccc|} \hline
$\Sigma(2N^2)$ & & $Q_N$ \\ \hline \hline
${\bf 2}_{m+k,m}$ & $\rightarrow$ & ${\bf 2}_{k}$ \\
${\bf 2}_{m+N/2,m}$ & $\rightarrow$ & ${\bf 1}_{+-} + {\bf 1}_{-+}$ \\
${\bf 1}_{+m}+ {\bf 1}_{-m}$  & $\rightarrow$ & ${\bf 1}_{--}+ {\bf 1}_{++}$ \\
 \hline
\end{tabular}
\end{center}
\caption{Decomposition of $\Sigma(2N^2)$ representations to 
$Q_N$ representations.}
\label{tab:QN}
\end{table}

\section{$T^2/Z_3$ orbifold}

Here, we consider the $T^2/Z_3$ orbifold, 
which has three fixed points.
We denote fields localized at these three fixed points 
by $\phi_i$ $(i=1,2,3)$.

\subsection{$S_3$}

The $T^2/Z_3$ orbifold has a large geometrical symmetry compared with 
the symmetry of $S^1/Z_2$.
First of all, there is a cyclic permutation symmetry among 
three fixed points.
Such a cyclic permutation symmetry acts on 
three localized fields as 
\begin{eqnarray}\label{eq:B-Z3}
B= \left(
\begin{array}{ccc}
0 & 1 & 0 \\
0 & 0 & 1 \\
1 & 0 & 0
\end{array}
\right),
\end{eqnarray}
on the basis $(\phi_1,\phi_2,\phi_3)^T$.
Obviously, this is the generator of the $Z_3$ symmetry, 
i.e. $B^3=I$.
In addition, the $T^2/Z_3$ orbifold has a reflection symmetry, 
where we exchange two fixed points each other with fixing 
the other fixed point.
One of such reflections can be represented by 
\begin{eqnarray}\label{eq:C-Z3}
C= \left(
\begin{array}{ccc}
1 & 0 & 0 \\
0 & 0 & 1 \\
0 & 1 & 0
\end{array}
\right),
\end{eqnarray}
on the basis $(\phi_1,\phi_2,\phi_3)^T$.
Being $C^2=I$ and $CBC=B^{-1}$,
the closed algebra of $B$ and $C$ correspond to 
$S_3$.
Thus, the $S_3$ flavor symmetry can be realized on 
the $T^2/Z_3$ orbifold.
The fields localized on three fixed points 
correspond to the combination of a singlet ${\bf 1}_+$
and a doublet ${\bf 2}$ of $S_3$.
Alternatively, we may assume that 
the generator $C$ is represented by 
\begin{eqnarray}\label{eq:C-Z3'}
C= -\left(
\begin{array}{ccc}
1 & 0 & 0 \\
0 & 0 & 1 \\
0 & 1 & 0
\end{array}
\right),
\end{eqnarray}
on the basis $(\phi_1,\phi_2,\phi_3)^T$.
In this case, the flavor symmetry is the same as $S_3$, 
but the fields localized on three fixed points 
correspond to the combination of ${\bf 1}_-$
and a doublet ${\bf 2}$ of $S_3$.

\subsection{$A_4$ and $\Delta(3N^2)$}

Similar to the previous section, here, we assume that 
the localized fields $\phi_i$ $(i=1,2,3)$ have 
$Z_N$ charges, $q_i$, under an additional $Z_N$ symmetry.
Then, such a $Z_N$ symmetry is represented as 
\begin{eqnarray}\label{eq:generic-A-Z3}
A= \left(
\begin{array}{ccc}
\rho^{q_1} & 0 & 0\\
0 & \rho^{q_2}& 0 \\
0 & 0 & \rho^{q_3}
\end{array}
\right),
\end{eqnarray}
on the basis $(\phi_1,\phi_2,\phi_3)^T$,
where $\rho = \exp(2\pi i/N)$.
For example, the $Z_3$ symmetry with $q_1=0, q_2=1$
and $q_3= 2$ can be realized in heterotic string models 
on the $T^2/Z_3$ orbifold.
Here, we consider more generic values of 
$Z_N$ charges $q_i$.

First, let us consider the model with $q_1=-q_3=1$ and 
$q_2=0$, i.e. 
\begin{eqnarray}\label{eq:A-Delta3N}
A= \left(
\begin{array}{ccc}
\rho & 0 & 0\\
0 & 1& 0 \\
0 & 0 & \rho^{-1}
\end{array}
\right),
\end{eqnarray}
on the basis $(\phi_1,\phi_2,\phi_3)^T$,
where $\rho =  \exp (2 \pi i /N)$.
For the moment, we assume that low-energy effective field theory 
has the cyclic permutation symmetry corresponding to $B$, but not 
the $Z_2$ reflection symmetry corresponding to $C$.
Hence, we study the closed algebra including $B$ and $A$ 
for Eq.~(\ref{eq:A-Delta3N}).
For example, we find $B^{-1}AB=A'$, where 
\begin{eqnarray}\label{eq:A'-Delta3N}
A'= \left(
\begin{array}{ccc}
\rho^{-1} & 0 & 0\\
0 & \rho& 0 \\
0 & 0 & 1
\end{array}
\right).
\end{eqnarray}
Indeed, their closed algebra corresponds to 
$\Delta(3N^2)$, which is isomorphic to 
$(Z_N \times Z_N') \rtimes Z_3$ \cite{Luhn:2007uq,Ishimori:2010au}.
All of $\Delta(3N^2)$ elements are written as 
$B^kA^mA'^n$ for $k=0,1,2$ and $m,n=0,1,\cdots,N-1$, 
and explicitly they are represented by 
\begin{eqnarray}
\left(
\begin{array}{ccc}
\rho^m & 0 & 0\\
0 & \rho^n & 0 \\
0 & 0 & \rho^{-m-n}
\end{array}
\right), \qquad 
\left(
\begin{array}{ccc}
0 & \rho^m & 0 \\
0 & 0 & \rho^n \\
 \rho^{-m-n} & 0 & 0 
\end{array}
\right), \qquad 
\left(
\begin{array}{ccc}
0 & 0 & \rho^m \\
\rho^n & 0 & 0\\
0 & \rho^{-m-n} & 0
\end{array}
\right), \qquad 
\end{eqnarray}
where $m,n=0,1,\cdots,N-1$.

Let us study representations of localized fields under 
$\Delta(3N^2)$ with smaller values of $N$.
The smallest and non-trivial model corresponds to 
$N=2$ and we obtain $\Delta(12)$, which is isomorphic to 
$A_4$.
In this model, we assume that $A$ in Eq.~(\ref{eq:A-Delta3N}) 
generates the $Z_2$ symmetry.
The fields localized on three fixed points have 
the $Z_2$ charges, $(q_1,q_2,q_3)=(1,0,1)$.
They correspond to an $A_4$ triplet ${\bf 3}$.
In addition, there can appear the localized fields, all of which 
have $Z_2$-even charges, $(q_1,q_2,q_3)=(0,0,0)$.
On those localized fields, the generator $A$ is represented 
trivially as $A=I$.
Thus, we can use the field basis diagonalizing the generator $B$,
i.e. 
\begin{eqnarray}\label{eq:A4-singlets}
\phi_1+\omega^n\phi_2+\omega^{2n}\phi_3,
\end{eqnarray}
for $n=0,1,2$ and $\omega = e^{2 \pi i/3}$.
In this basis, the generator $B$ is represented by $B=\omega^n$ 
for $n=0,1,2$.
These correspond to three types of $A_4$ singlets, 
${\bf 1}$, ${\bf 1}'$ and ${\bf 1}''$.
Note that singlets appear only in the combination, 
${\bf 1} + {\bf 1}' + {\bf 1}''$ for the localized fields.
On the other hand, a trivial singlet ${\bf 1}$
can appear as bulk modes.
These results are shown in Table \ref{tab:A4}.
For example, in the $A_4$ lepton model by Altarelli and Feruglio 
\cite{Altarelli:2005yp},  
three families of left-handed and right-handed charged leptons 
are assigned with ${\bf 3}$ and ${\bf 1} + {\bf 1}' + {\bf 1}''$, 
respectively, while Higgs fields correspond to trivial singlets ${\bf
  1}$.
In addition, all of flavon fields correspond to 
trivial singlets ${\bf 1}$ and triplets ${\bf 3}$.
Such representations can be realized on the $T^2/Z_3$ orbifold 
model with proper $Z_2$ charges.

\begin{table}[t]
\begin{center}
\begin{tabular}{|c|c|} \hline
$Z_2$ charges of $(\phi_1,\phi_2,\phi_3)$ & Representation of $A_4$ \\ \hline \hline
$(1,0,1)$ & ${\bf 3}$ \\    \hline 
$(0,0,0)$ & ${\bf 1}+ {\bf 1}'+  {\bf 1}''$ \\ \hline
\end{tabular}
\end{center}
\caption{$A_4$ representations of the fields localized on the
  $T^2/Z_3$ orbifold.}
\label{tab:A4}
\end{table}

Next, we consider the model with $N=3$, where 
we obtain $\Delta(27)$.\footnote{The $\Delta(27)$ symmetry 
can be realized in magnetized/intersecting brane models 
\cite{Abe:2009vi}.}
In this model, we assume that $A$ in Eq.~(\ref{eq:A-Delta3N}) 
generates the $Z_3$ symmetry.
Then, the fields localized on three fixed points have 
the $Z_3$ charges, $(q_1,q_2,q_3)=(1,0,2)$, and 
they correspond to a triplet ${\bf 3}_1$.
In addition, there are also the localized fields 
with $Z_3$ charges, $(q_1,q_2,q_3)=(2,0,1)$ and 
they correspond to another triplet ${\bf 3}_2$.
Moreover, there are also the localized fields 
with $Z_3$ charges, $(q_1,q_2,q_3)=(m,m,m)$, for 
$m=0,1,2$.
On these fields, the generator $A$ is represented by 
$A=\omega^m I$ for $m=0,1,2$.
For these fields, we can use the field basis 
diagonalizing the generator $B$, i.e. 
$\phi_1+\omega^n\phi_2+\omega^{2n}\phi_3$ for $n=0,1,2$.
In this basis, the generator $B$ is represented by $B=\omega^n$ 
for $n=0,1,2$.
These correspond to nine singlets of $\Delta(27)$, 
${\bf 1}_{nm}$, on which the generators $A$ and $B$ are 
represented by $A=\omega^m$ and $B=\omega^n$.
Note that only the combinations 
${\bf 1}_{0m}+{\bf 1}_{1m}+{\bf 1}_{2m}$ appear 
on three fixed points.
These results are shown in Table \ref{tab:Delta27}.

\begin{table}[t]
\begin{center}
\begin{tabular}{|c|c|} \hline
$Z_3$ charges of $(\phi_1,\phi_2,\phi_3)$ & Representation of $\Delta(27)$ \\ \hline \hline
$(1,0,2)$ & ${\bf 3}_1$ \\    \hline 
$(2,0,1)$ & ${\bf 3}_2$ \\    \hline 
$(0,0,0)$ & ${\bf 1}_{00}+ {\bf 1}_{10}+  {\bf 1}_{20}$ \\ \hline
$(1,1,1)$ & ${\bf 1}_{01}+ {\bf 1}_{11}+  {\bf 1}_{21}$ \\ \hline
$(2,2,2)$ & ${\bf 1}_{02}+ {\bf 1}_{12}+  {\bf 1}_{22}$ \\ \hline
\end{tabular}
\end{center}
\caption{$\Delta(27)$ representations of the fields localized on the
  $T^2/Z_3$ orbifold.}
\label{tab:Delta27}
\end{table}

Similarly, we can study the $T^2/Z_3$ orbifold models 
with $Z_N$ charges for $N>3$.
The $\Delta(3N^2)$ flavor symmetry can be realized 
and certain representations can be obtained on fixed points such as 
all of triplets and certain combinations of singlets.
Furthermore, subgroups of $\Delta(3N^2)$ can also be realized.
For example, the $T_7$ flavor symmetry, which is isomorphic to 
$Z_7 \rtimes Z_3$ \cite{Ma:2007ia,Hagedorn:2008bc,Ishimori:2010au}, 
can be realized by the generator 
$A= {\rm diag}(\rho,\rho^2,\rho^4)$ with $\rho=\exp (2\pi i/7)$ 
and the generator $B$.
Other subgroups of $\Delta(3N^2)$ can also be realized.

\subsection{$S_4$ and $\Delta(6N^2)$}

Here, let us study the model, where the geometrical symmetry 
includes both $B$ of Eq.~(\ref{eq:B-Z3}) and $C$ of Eq.~(\ref{eq:C-Z3}) 
and fields also have $Z_N$ charges 
corresponding to $A$ of Eq.~(\ref{eq:A-Delta3N}).
Their closed algebra corresponds to $\Delta(6N^2)$, 
which is isomorphic to 
$(Z_N \times Z_N) \rtimes S_3$ \cite{Escobar:2008vc,Ishimori:2010au}.
All of $\Delta(6N^2)$ elements are written as 
\begin{eqnarray}
B^kC^\ell A^m A'^n,
\end{eqnarray}
for $k=0,1,2$, $\ell = 0,1$ 
and $m,n=0,1,\cdots,N-1$.

Let us study representations of localized 
fields under $\Delta(6N^2)$ with smaller values 
of $N$ as the previous section.
Here, we start with $N=2$ by using the generator $A$ in 
 Eq.~(\ref{eq:A-Delta3N}) as the $Z_2$ symmetry.
Then, we obtain $\Delta(24)$, which is isomorphic to 
$S_4$.
The fields localized on three fixed points have 
$Z_2$ charges $(1,0,1)$.
They correspond to a triplet ${\bf 3}$ of $S_4$.
The localized fields with $Z_2$ charges 
$(0,0,0)$ correspond to 
the combinations, ${\bf 1} + {\bf 2}$ of $S_4$.
When we use the generator $C$ of Eq.~(\ref{eq:C-Z3'})
instead of Eq.~(\ref{eq:C-Z3}), then the three localized fields 
with $Z_2$ charges $(1,0,1)$ and $(0,0,0)$ correspond to 
another triplet ${\bf 3}'$ and the combination ${\bf 1}' + {\bf 2}$, 
respectively.
These results are shown in Table \ref{tab:S4}.
The bulk fields would appear as a trivial singlets ${\bf 1}$.
Such a singlet may have a mass term with the trivial singlet 
of the combinations of the three localized fields, 
${\bf 1} + {\bf 2}$.
Then, below such a mass scale, the doublets ${\bf 2}$ 
remain as light modes.
Furthermore, the doublets between the combinations, 
${\bf 1} + {\bf 2}$ and 
${\bf 1}'+ {\bf 2}$ may have mass terms.
Below such a mass scale, only the non-trivial singlet ${\bf 1}'$ 
remains as a light mode.
Hence, any combinations of ${\bf 1}$, ${\bf 1}'$, ${\bf 2}$, ${\bf
3}$ and ${\bf 3}'$
can appear on the $T^2/Z_3$ orbifold.
Then, we can realize any combinations of all $S_4$ representations.
Therefore, we can realize any $S_4$ models on the $T^2/Z_3$ orbifold.

\begin{table}[t]
\begin{center}
\begin{tabular}{|c|c|c|} \hline
$Z_2$ charges of $(\phi_1,\phi_2,\phi_3)$ & Representation of $S_4$ 
& Representation of $S_4$ \\ 
  & with $C$ in Eq.~(\ref{eq:C-Z3}) &  with $C$ in Eq.~(\ref{eq:C-Z3'}) \\
\hline \hline
$(1,0,1)$ & ${\bf 3}$ & ${\bf 3}'$ \\    \hline 
$(0,0,0)$ & ${\bf 1}+ {\bf 2}$ & ${\bf 1}'+ {\bf 2}$ \\ \hline
\end{tabular}
\end{center}
\caption{$S_4$ representations of the fields localized on the
  $T^2/Z_3$ orbifold.}
\label{tab:S4}
\end{table}

Next, we consider the model with $N=3$.
In this model, the generator $A$ can be represented by 
$A = {\rm diag}(\omega,1,\omega^2)$, and the full flavor symmetry 
corresponds to $\Delta(54)$.
Indeed, this flavor symmetry can be realized in heterotic 
string theory on the $T^2/Z_3$ orbifold \cite{Kobayashi:2006wq}.
For the generator $C$ in Eq.~(\ref{eq:C-Z3}), 
the localized fields with the $Z_3$ charges 
$(0,k,2k)$ $(k=1,2)$ on the three fixed points 
correspond to a triplet ${\bf 3}_{1(k)}$ 
of $\Delta(54)$.
In addition, the localized fields $(\phi_1,\phi_2,\phi_3)$ 
with the trivial $Z_3$ charge $q_i=0$ for $i=1,2,3$ 
correspond to a trivial singlet ${\bf 1}_+$ and a doublet ${\bf 2}_1$. 
Here, the singlet is the linear combination, 
$\phi_1+\phi_2+\phi_3$, while the doublet 
corresponds to 
\begin{eqnarray}
(\phi_1+\omega \phi_2+ \omega^2\phi_3,~~  
\phi_1+\omega^2 \phi_2+ \omega \phi_3).
\end{eqnarray}
When the generator $C$ is represented by Eq.~(\ref{eq:C-Z3'}), 
the localized fields with the $Z_3$ charges 
$(0,k,2k)$ $(k=1,2)$ on the three fixed points 
correspond to a triplet ${\bf 3}_{2(k)}$, 
and  the localized fields $(\phi_1,\phi_2,\phi_3)$ 
with the trivial $Z_3$ charge $q_i=0$ 
correspond to a non-trivial singlet ${\bf 1}_-$ and a doublet ${\bf 2}_1$. 
Similarly, we can study the $T^2/Z_3$ orbifold models with $Z_N$ 
charges for $N > 3$, where the $\Delta(6N^2)$ flavor symmetry can be realized.

\subsection{Larger symmetries}

We can realize larger flavor symmetries by using 
generic values of $Z_N$ charges, i.e.  the generator $A$ of 
Eq.~(\ref{eq:generic-A-Z3}) with generic values of $q_1, q_2$ 
and $q_3$.
For example, the combination between such a generator $A$ and 
the generator $B$ would lead to the flavor symmetry, 
$\Sigma(3N^3)$.
All of the $\Sigma(3N^3)$ elements are written as 
\begin{eqnarray}\label{eq:Sigma3N3}
\left(
\begin{array}{ccc}
\rho^\ell & 0 & 0 \\
0 & \rho^m & 0 \\
0 & 0 & \rho^n
\end{array}
\right), \qquad 
\left(
\begin{array}{ccc}
0 & \rho^m & 0  \\
0 & 0 & \rho^n \\
\rho^\ell & 0 & 0
\end{array}
\right), \qquad 
\left(
\begin{array}{ccc}
0 & 0 & \rho^n \\
\rho^\ell & 0 & 0  \\
0 & \rho^m & 0 
\end{array}
\right).
\end{eqnarray}
Here, the group for $N=2$ is isomorphic to 
$Z_2 \times \Delta(12)$ and $Z_2 \times A_4$.
Thus, the non-trivial group with the smallest $N$ is 
$\Sigma(81)$ \cite{Hagedorn:2008bc,Ishimori:2010au}.

Furthermore,  we can include the generator $C$ of Eq.~(\ref{eq:C-Z3}) 
in the above algebra.
Then, the $Z_3$ part generated by $B$ is replaced by 
$S_3$ generated by $B$ and $C$, and the total symmetry becomes larger.
All elements are written by 
\begin{eqnarray}
\left(
\begin{array}{ccc}
\rho^\ell & 0 & 0 \\
0 & 0 & \rho^m \\
0 &  \rho^n & 0
\end{array}
\right), \qquad 
\left(
\begin{array}{ccc}
0 & 0 & \rho^m  \\
0 &  \rho^n &0 \\
\rho^\ell & 0 & 0
\end{array}
\right), \qquad 
\left(
\begin{array}{ccc}
0 &  \rho^n& 0 \\
\rho^\ell & 0 & 0  \\
0 & 0 & \rho^m  
\end{array}
\right),
\end{eqnarray}
in addition to the above $\Sigma(3N^3)$ elements (\ref{eq:Sigma3N3}).

\section{Conclusion}

We have studied non-Abelian discrete flavor symmetries 
on the $S^1/Z_2$ and $T^2/Z_3$ orbifolds.
We have introduced $Z_N$ charges for fields localized 
on orbifold fixed points and classified the possible symmetries.
By combining $Z_N$ symmetries and geometrical symmetries 
of the orbifolds, we can realize $D_N$, $\Sigma(2N^2)$, $\Delta(3N^2)$ and 
$\Delta(6N^2)$ including $S_3$, $A_4$, $S_4$.
Their subgroups are also obtained such as $Q_N$, $T_7$, etc.
Furthermore, larger flavor symmetries such as $\Sigma(3N^3)$ are also possible.
Thus, our models provide with geometrical setups 
of models using these flavor symmetries.
In general, certain combinations of representations are allowed to appear.
However, any combinations of representations can appear, e.g. for
$S_3$ and $S_4$ in low-energy effective theory by using certain mass terms.
Although we have constraints on combinations of representations in
$A_4$ models, our setups could fit assignments of the model by 
Altarelli and Feruglio and other models. 
Therefore, our orbifold setups could realize several interesting
models.

To derive realistic quark/lepton masses and mixing angles, 
we have to break non-Abelian discrete flavor symmetries 
along certain directions.
Orbifolds are also useful to break flavor symmetries 
by boundary conditions
\cite{Haba:2006dz,Kobayashi:2008ih,Seidl:2008yf,
Watanabe:2010cy,Burrows:2010wz}.
Furthermore, anomalies of non-Abelian discrete symmetries 
would also lead to an important constraint 
\cite{Araki:2007zza,Araki:2008ek,Luhn:2008sa,Ishimori:2010au}.
We leave these studies as future work.

\subsection*{Acknowledgement}

H.~A. is supported in part by the Waseda University Grant 
for Special Research Projects No. 2010B-185.
K.-S.~C., T.~K. and H.~O. are supported in part by the Grant-in-Aid for 
Scientific Research No.~20$\cdot$08326, No.~20540266 and
No.~21$\cdot$897 from the 
Ministry of Education, Culture, Sports, Science and Technology of Japan.
T.~K. is also supported in part by the Grant-in-Aid for the Global COE 
Program "The Next Generation of Physics, Spun from Universality and 
Emergence" from the Ministry of Education, Culture,Sports, Science and 
Technology of Japan.

\end{document}